\documentclass[onecolumn,notitlepage,letterpaper,superscriptaddress,nofootinbib,longbibliography,prd]{revtex4-2}

\usepackage[english]{babel}
\usepackage[T1]{fontenc}
\usepackage[utf8]{inputenc}

\usepackage{amsmath,amssymb,amsfonts,accents}
\usepackage{graphicx}
\usepackage{xcolor}
\usepackage{float}
\usepackage{comment}
\usepackage{mathrsfs}
\usepackage[normalem]{ulem}

\usepackage{caption}
\usepackage[dvipsnames]{xcolor}
\usepackage[
  pdftoolbar=true,
  pdfstartview=FitH,
  pdfmenubar=true
]{hyperref}
\usepackage{xurl}

\hypersetup{
  colorlinks=true,
  linkcolor=blue,
  filecolor=magenta,
  citecolor=blue,
  urlcolor=blue
}

\usepackage{orcidlink}

\newcommand{\beq}{\begin{equation}}
\newcommand{\eeq}{\end{equation}}

\begin{document}

\title{Frame invariant diffusive formulation of scalar-tensor gravity}

\author{Laur J\"arv\orcidlink{0000-0001-8879-3890}}
\email{laur.jarv@ut.ee}
\affiliation{Institute of Physics, University of Tartu, W.\ Ostwaldi 1, 50411 Tartu, Estonia}

\author{Sotirios Karamitsos\orcidlink{0000-0002-1773-249X}}
\email{sotirios.karamitsos@ut.ee}
\affiliation{Institute of Physics, University of Tartu, W.\ Ostwaldi 1, 50411 Tartu, Estonia}

\date{\today}

 \begin{abstract}
Thermodynamics provides a useful interpretation of scalar-tensor gravity, in which the effective imperfect fluid admitted by the nonminimal coupling features a temperature that is associated with the departure from general relativity. However, in this construction, certain thermodynamical quantities are defined with respect to a particular conformal frame. In the present work, we show that the originally proposed effective temperature assigned to nonminimally coupled scalar field theories is not frame invariant, and can thus be arbitrarily tuned by a change of frame. This raises the question of whether temperature can be viewed as an intrinsic property of a scalar-tensor theory rather than a particular representation of it. Working instead with the frame invariant formulation of scalar-tensor gravity, we find that the frame invariant effective fluid is perfect with identically vanishing temperature. The departure from general relativity is then governed not by temperature, but rather by a frame invariant chemical potential, similar to minimal theories. Therefore, general relativity can be interpreted as a state of diffusive equilibrium for any scalar-tensor theory, regardless of whether it is minimal or nonminimal.
\end{abstract}

\maketitle
 
 \section{Introduction}

There are several thermodynamical results borne out of the edifice of general relativity (GR), especially in the study of black holes, where the laws of black hole thermodynamics can be derived from the Einstein field equations \cite{Wald:1999vt,Padmanabhan:2009vy,Wall:2018ydq,Witten:2024upt}. However, it is possible to reason in the opposite direction and derive the field equations of GR from thermodynamical considerations of entropy and horizon area \cite{Jacobson:1995ab}. This is a surprising result\footnote{Or perhaps not so surprising, given Einstein's fondness for thermodynamics, famously stating that \emph{``[thermodynamics] is the only physical theory of universal content which I am convinced will never be overthrown, within the framework of applicability of its basic concepts.''}} that may favor an emergent treatment of gravity. Furthermore, the thermodynamical treatment of black holes and gravity is now known to extend to theories beyond GR \cite{Sarkar:2019xfd}. In this view, GR corresponds to equilibrium thermodynamics, while departures from GR correspond to dissipative thermodynamics~\cite{Eling:2006aw}.

The various mathematical conundrums, as well as the multifaceted puzzles of contemporary cosmology offer strong motivations to probe the extensions and modifications of Einstein's theory of GR \cite{Heisenberg:2018vsk,CosmoVerseNetwork:2025alb,CANTATA:2021asi}.
One of the most enduring classes of modified gravity theories is scalar-tensor gravity \cite{Jordan1955,Fierz:1956zz,Brans:1961sx}. In these theories, a massive scalar degree of freedom is introduced in addition to the usual spin-2 degree of freedom to mediate the gravitational interaction.  Scalar degrees of freedom arise naturally both in high-energy physics (often as low-energy EFTs that appear as remnants of more fundamental physics) as well as in various modified theories of gravity, such as $F(R)$ gravity or metric-affine gravity. The prototypical example of a scalar-tensor theory is perhaps that of Brans--Dicke theories~\cite{Brans:1961sx}, in which the scalar field takes on the role of the effective gravitational coupling. Brans--Dicke theories are further indexed by the Brans--Dicke coupling $\omega$, which may also be promoted to a function $\omega(\phi)$, seemingly capturing a wider class of theories. More sophisticated scalar-tensor theories of gravity have been proposed, including Horndeski, DHOST, and beyond, although gravitational wave observations have drastically constrained the allowed range of these theories \cite{Kobayashi:2019hrl}.  

It is well understood that modified theories of gravity can be studied in terms of effective energy-momentum tensors \cite{Capozziello:2011et,Capozziello:2013vna}. Furthermore, effective energy-momentum tensors for scalar-tensor theories can be interpreted as corresponding to an imperfect fluid \cite{Pimentel:1989bm,Pimentel:2016jlm}, described by standard thermodynamical quantities such as heat flux, entropy density, and temperature.  This has led to renewed interest in the literature in the deformation of modified gravity towards the equilibrium of GR in terms of thermodynamical quantities \cite{Faraoni:2021lfc,Giusti:2021sku, Faraoni:2021jri,Giardino:2022sdv,Faraoni:2022gry, Gallerani:2024gdy,Gallerani:2025myd,Faraoni:2025alq}, and it has been demonstrated that dissipative relativistic thermodynamics, as studied by Eckart \cite{Eckart:1940te}, is a very suitable toolset to apply to modified gravity theories.

Scalar-tensor theories can be naturally cast in different frames depending on whether the nonminimal coupling between the scalar field and curvature appears explicitly (Jordan frame) or not (Einstein frame). It is possible to transform between frames by means of conformal transformations (in this context, a field-dependent rescaling of the metric) and an internal reparametrization of the fields. The discussion of whether one frame is privileged over others is known as the \emph{frame problem}, and it has been the subject of much debate in the literature \cite{Faraoni:1999hp,
    Capozziello:2006dj, 
    Catena:2006bd, 
    Faraoni:2006fx, 
    Capozziello:2010sc, 
    Weenink:2010rr,
    Gong:2011qe, 
    Quiros:2011iv, 
    Kubota:2011re, 
    White:2012ya, 
    Quiros:2012rnn, 
    Calmet:2012eq, 
    Prokopec:2013zya, 
    White:2013ufa, 
    Ruf:2017xon, Falls:2018olk}. 
Claims of frame-dependence appear in the context of certain quantities like the number of e-folds \cite{Karam:2019dlv} or reheating quantities \cite{Dorsch:2026ref}, or at the quantum level  when radiative corrections are involved \cite{Steinwachs:2013tr,Kamenshchik:2014waa,Domenech:2015qoa}, although there are claims of quantum equivalence of different conformal frames as well \cite{Gasperini:1993hu, Karamitsos:2017elm, Ohta:2017trn}. Still, the broad consensus is that the choice of frame holds no physical meaning at the classical level  \cite{Flanagan:2004bz,Chiba:2013mha,Postma:2014vaa,Jarv:2014hma, Burns:2016ric}.
A way we can demonstrate this is by considering observable quantities and showing that they are frame invariant. There is an apt analogy with dimensionful and dimensionless quantities, already identified by Dicke~\cite{Dicke:1961gz}: the numerical value of dimensionful quantities transforms arbitrarily depending on the system of units used, whereas the numerical value of dimensionless quantities remains unchanged. The discussion of whether dimensions are physical or ``man-made'' is more suited to ontology rather than physics. However, it is known that physically meaningful relations between quantities \emph{can} always be written in a nondimensionalized form (see the Buckingham-$\pi$ theorem).        

All this brings us to the thermal invariants of scalar-tensor gravity. Despite the robust results derived from applying dissipative fluid thermodynamics to the study of scalar-tensor theories, it appears that many quantities that arise in the application of this formalism are frame-dependent. This is usually done by working in the Brans--Dicke representation. However, it is possible to recast any scalar-tensor theory in a Brans--Dicke-like form, and in fact, there are multiple equivalent ways to do so. This means that a thermodynamical quantity expressed by working in the Brans--Dicke picture must be frame-dependent, and can be tuned by appropriately changing between equivalent Brans--Dicke representations. As such, even though the temperature of scalar-tensor theories may contain useful information about the relaxation of a particular representation of scalar-tensor theories towards GR, its definition depends heavily on the frame in which the theory is expressed, which makes it less useful as a means of distinguishing between theories.  

Therefore, we ask whether it is possible to define thermodynamical quantities in a frame invariant manner. To answer this question, we first note that when we write down a theory, we must choose a frame in practice. The choice is arbitrary, and only influenced by ease of presentation and calculation: there is no preferred conformal frame. However,  it is always possible to write a scalar-tensor theory in a manifestly frame invariant manner such that every quantity that appears in the action is frame invariant \cite{Jarv:2014hma,Jarv:2015kga,Jarv:2016sow}. This allows us to isolate the features of an action that depend on the way that it is parametrized, and those that are independent of its representation. As such, the invariant functions are more ``fundamental'', in the sense that they are features of the underlying theory and not its representation. In this sense, frame covariance is analogous to general covariance and the distinction between relative and absolute quantities: frame invariant quantities describe a theory regardless of representation, whereas frame-dependent quantities only make sense once a frame is specified. Our aim in this work is to define a set of frame invariant thermodynamical and hydrodynamical quantities from the ground up, therefore encoding the relaxation of scalar-tensor theories to the GR equilibrium in a completely frame-agnostic manner.
 
The structure of the paper is as follows: in Section~\ref{sec:firstorder}, we present an overview of the first-order thermodynamical description admitted by Brans--Dicke theories as well as the diffusive picture that emerges when considering the effective fluid description of minimally coupled theories. We review the frame-covariant and frame invariant approach to scalar-tensor theories in Section~\ref{sec:invariantscalartensor} before demonstrating in Section~\ref{sec:conformallydependentthermo} that the usual definition of temperature is frame-dependent, both in Brans--Dicke theories as well as more general scalar-tensor theories. In Section~\ref{sec:frameinvariantequilibrium}, we show that a frame invariant treatment of matter-free scalar-tensor theories admits a non-dissipative fluid with vanishing temperature and finite chemical potential. Finally, we discuss multifield scalar theories and the associated single chemical potential in Section~\ref{sec:multifield} before concluding in Section~\ref{sec:conclusions}.

Throughout the paper, we use the mostly minus convention for the metric tensor, and we work in natural units where $\hbar = G =c = 1$.
  
\section{First-order thermodynamics of Brans--Dicke theories}\label{sec:firstorder}

We begin by examining the well-known and very general class of Brans--Dicke theories as expressed in the Jordan frame, which are specified by the following scalar-tensor action:
\begin{equation}
S_{\text{BD}} = \frac{1}{2}\int d^{4}x \sqrt{-g} 
\left[ 
\phi R - \frac{\omega(\phi)}{\phi} \nabla^\mu\phi \nabla_\mu\phi - V(\phi) 
\right] 
+ S^{(\rm m)},
\end{equation} 
where $R$ is the scalar curvature and $\omega(\phi)$ is the Brans--Dicke coupling function, promoted from the usual Brans--Dicke coupling constant $\omega$. In this action, we assume that the matter is coupled to the metric $g_{\mu\nu}$ as $S^{(\text{m})}  [g_{\mu\nu}, \chi ]$, where $\chi$ collectively denotes the matter fields.

With the usual variational methods, it is possible to write down the equations of motion as follows:
\begin{align}
  G_{\mu\nu} &=  \frac{T^{\rm (m)}_{\mu\nu}}{\phi}+ \frac{\omega}{\phi^2} \left[ \nabla_\mu \phi \nabla_\nu \phi - \frac{g_{\mu\nu}}{2} (\nabla \phi)^2   \right] +  \frac{1}{\phi}(\nabla_\mu \nabla_\nu \phi - g_{\mu\nu} \Box \phi) - \frac{g_{\mu\nu}}{\phi} \frac{V}{2},
\\
\Box \phi &= \frac{1}{2\omega+3}\left[ T^{\rm (m)} + \phi V' - 2V  - \omega' (\nabla\phi)^2\right],
\label{eq: scalar field equation}
\end{align}
where $G_{\mu\nu} \equiv R_{\mu\nu} - \frac{g_{\mu\nu}R}{2}$ is as usual the Einstein tensor, $\Box \equiv g^{\mu\nu} \nabla_\mu \nabla_\nu$, and primes indicate differentiation with respect to the scalar field $\phi$ (we have suppressed arguments of $\phi$ for readability). We denote the trace of the matter energy-momentum tensor as $T^{\rm (m)} \equiv g^{\mu\nu} T_{\mu\nu}^{\rm (m)}$.

Separating the non-gravitational terms from the contributions arising due to the presence of the field is useful, because it allows us to define an \emph{effective} energy-momentum tensor for $\phi$:
\begin{align}\label{fluidemtensor}
 T^{(\phi)}_{\mu\nu} =   \frac{\omega}{\phi^2} \left[ \nabla_\mu \phi \nabla_\nu \phi - \frac{g_{\mu\nu}}{2} (\nabla \phi)^2   \right] + \frac{1}{\phi}(\nabla_\mu \nabla_\nu \phi - g_{\mu\nu} \Box \phi) - \frac{g_{\mu\nu}}{\phi}\frac{V}{2}.
\end{align}
With this definition, we can rewrite the Einstein field equations as
\begin{align}
 G_{\mu\nu} &=  \frac{T^{\rm (m)}_{\mu\nu}}{\phi}+     T^{(\phi)}_{\mu\nu}.
\end{align}
We note that the prefactor of $\phi$ is absorbed in the definition of the effective energy-momentum tensor.
 
The effective stress-energy tensor defined in \eqref{fluidemtensor} corresponds, in fact, to that of an imperfect fluid \cite{Faraoni:2018qdr}. The only assumption necessary for the imperfect fluid description is that the gradient of the scalar field is timelike and future-oriented \cite{Faraoni:2021lfc, Giardino:2023ygc}. The existence of a future-oriented velocity $u_\mu$ allows us to foliate spacetime in three-dimensional sheets that correspond to the space of the observers that are comoving with the fluid. Then spacetime is endowed with an induced metric
\begin{align}
h_{\mu\nu} \equiv  g_{\mu\nu} + u_\mu u_\nu.
\end{align}
The induced metric may be used to project down to the comoving space, therefore allowing for the definition of the spatial projection of the velocity gradient $\nabla_\rho u_\sigma$:
\begin{align}
V_{\mu\nu} &\equiv h_\mu{}^\rho h_\nu{}^\sigma \nabla_\rho u_\sigma.
\end{align}
This can be decomposed into a symmetric part (the expansion tensor $\theta_{\mu\nu}$) and an antisymmetric part (the vorticity $\omega_{\mu\nu}$):
\begin{align}
\theta_{\mu\nu} &\equiv h_\mu{}^\rho h_\nu{}^\sigma \nabla_{(\rho} u_{\sigma)},
 \\
\omega_{\mu\nu} &\equiv h_\mu{}^\rho h_\nu{}^\sigma \nabla_{[\rho} u_{\sigma]}.
\end{align}
The expansion tensor can be further decomposed in terms of the expansion scalar $\theta$ and the symmetric traceless shear tensor $\sigma_{\mu\nu}$:
\begin{align}\label{exptensdec}
\theta_{\mu\nu} = \sigma_{\mu\nu} + \frac{\theta}{3} h_{\mu\nu} .
\end{align}
We can also define the shear scalar and the vorticity scalar:
\begin{align}
\sigma \equiv  \frac{1}{2} \sigma_{\mu\nu}\sigma^{\mu\nu}, \qquad
\omega \equiv  \frac{1}{2} \omega_{\mu\nu}\omega^{\mu\nu}
\end{align}
It is also helpful to define the effective four-acceleration of the fluid as the rate of change of velocity along flow lines, 
\begin{align}\label{fouraccelerationdef}
a^\mu \equiv \frac{d u^\mu}{d\tau} = u^\nu \nabla_\nu u^\mu,
\end{align}
where $\tau$ is the proper time defined along flow lines. Analogously, we can also define the rate of change of any quantity~$Y$  along flow lines as $dY/d\tau\equiv u^\nu \nabla_\nu Y$.

Expanding the projection operators $h_\mu^{\ \rho} = \delta^\rho_\mu + u_\mu u^\rho$ and using $u^\nu \nabla_\mu u_\nu = 0$ (which follows from differentiating $u_\mu u^\mu = -1$), we can write
\begin{align}
\nabla_\mu u_\nu &=  \theta_{\mu\nu} + \omega_{\mu\nu} - u_\mu a_\nu
\\
& = \sigma_{\mu\nu} + \frac{\theta}{3} h_{\mu\nu} - u_\mu {a}_\nu,
\end{align}
where we have decomposed the expansion tensor as in \eqref{exptensdec} and used the fact that the vorticity vanishes because the velocity is a gradient.

So far, we have only written down purely kinematic quantities. In order to make a connection to thermodynamics, we must first arrive at an expression for mechanical quantities (such as energy and pressure). To do so, we look at the imperfect fluid decomposition, which can be written as
\begin{align}
T_{\mu\nu} = \rho u_\mu u_\nu + q_\mu u_\nu + u_\mu q_\nu + \Pi_{\mu\nu},
\end{align}
or equivalently as
\begin{align}
 T_{\mu\nu} = \rho u_\mu u_\nu + ph_{\mu\nu}+ q_\mu u_\nu + u_\mu q_\nu + \pi_{\mu\nu}.
\end{align}
With this decomposition, several quantities are defined. These are: the comoving effective energy density $\rho$, the heat flux $q_\mu$, the stress tensor $\Pi_{\mu\nu}$, the isotropic pressure~$p$, and the anisotropic stress $\pi_{\mu\nu}$ (which is simply the trace-free part of $\Pi_{\mu\nu}$). These quantities are given in terms of kinematic quantities and the energy-momentum tensor as follows:  
\begin{align}
\rho &= T_{\mu\nu} u^\mu u^\nu,
\\
\label{heatfluxdef}
q_\mu &= -T_{\rho \sigma} u^\rho h_\mu{}^\sigma,
\\
\Pi_{\mu\nu} &= T_{\rho\sigma} h_{\mu}{}^\rho h_\nu{}^\sigma,
\\
p &= \frac{h^{\mu\nu} }{3} T_{\mu\nu},
\\
\pi_{\mu\nu} &= \Pi_{\mu\nu} - p h_{\mu\nu}.
\end{align}
where we have used $p = \frac{g^{\mu\nu}}{3} \Pi_{\mu\nu} $ in writing $p$.

We have not yet made the link to non-equilibrium thermodynamics. To do so, we must consider the constitutive equations for dissipative quantities (``dissipative'' in the sense that they vanish in local equilibrium and are responsible for entropy production). These can be found in the formulation of first-order relativistic thermodynamics (studied by Eckart\footnote{Eckart's first-order thermodynamics is famously acausal and features generic instabilities \cite{HISCOCK1983466}, since disturbances in the medium propagate instantaneously. The Israel--Stewart approach \cite{ISRAEL1979341, ISRAEL1976310} introduces relaxation times in the medium to rectify this issue. However, first-order thermodynamics is still commonly used as a rough approximation, and is enough to establish a relation between thermodynamics and scalar-tensor theories.} \cite{Eckart:1940te}). The constitutive equations for the heat flux $q_\mu$ and the anisotropic stress tensor $\pi_{\mu\nu}$, along with the (bulk) viscous pressure $p_{\rm visc}$, are 
\begin{align}
p_{\rm visc} &= - \zeta \theta,
\\
\label{consteq2}
q_\mu &= -K (h_{\mu\nu} \nabla^\nu T + T {a}_\mu),
\\
\pi_{\mu\nu} &= -2\eta \sigma_{\mu\nu}.
\end{align}
The properties of the medium are encoded in the bulk viscosity~$\zeta$, the thermal conductivity $K$, and the shear viscosity~$\eta$, whereas~$T$ is the temperature.

In general, there is no apparent \emph{a priori} reason why the effective fluid description of a theory of gravity (admitting a particular stress-energy tensor) should follow a thermodynamical law. However, if the constitutive equations are formally satisfied by the effective fluid description admitted by the scalar-tensor theory, then said equations will effectively provide a \emph{thermodynamical interpretation} of the scalar-tensor theory.
Following Refs.\ \cite{Faraoni:2021lfc, Giardino:2023ygc, Faraoni:2025alq, Gallerani:2025myd}, we will now proceed to examine whether this is the case.

Working in the Eckart (relativistic) frame in which the velocity aligns with the particle current, we can construct the normalized effective fluid four-velocity for the scalar-tensor theory as
\begin{align}\label{fourvelocitydef}
u_\mu \equiv \frac{\nabla_\mu\phi}{\sqrt{-\nabla_\rho \phi \nabla^\rho \phi}}.
\end{align}
We are now ready to relate dissipative quantities to the scalar-tensor quantities. The most relevant equation is Fourier's law \eqref{consteq2}, for which we need to relate the expression for the four-acceleration $a_\mu$ as well as the kinematic expression for the heat flux $q_\mu$. The explicit expression for $a_\mu$, which we can find simply through \eqref{fourvelocitydef} and \eqref{fouraccelerationdef}, is:
\begin{align}\label{acceldef}
a_\mu =\frac{ (\nabla^\nu \phi)}{ (-\nabla^\rho \phi \nabla_\rho \phi)^{2}} 
\big[ 
(-\nabla^\rho \phi \nabla_\rho \phi) ( \nabla_\mu \nabla_\nu \phi) + (\nabla^\sigma \phi )(\nabla_\sigma \nabla_\nu \phi) (\nabla_\mu \phi)
\big],
\end{align}
Explicitly calculating the heat flux $q_\mu$ from \eqref{heatfluxdef} and the expression for the effective stress-energy tensor \eqref{fluidemtensor}, we find
\begin{align}
q_\mu = \frac{(\nabla^\rho \phi )(\nabla^\sigma \phi) }{(-\nabla^\nu \phi \nabla_\nu\phi)^{3/2}} 
\big[ 
(\nabla_\sigma \phi) (\nabla_\rho \nabla_\mu \phi) - (\nabla_\mu \phi) (\nabla_\rho \nabla_\sigma \phi) 
\big],
\end{align}
We therefore note that the two are related simply by
\begin{align}
q_{\mu}  = -\frac{\sqrt{-\nabla^\rho \phi \nabla_\rho \phi}}{\phi} {a}_{\mu}.
\end{align}
We may therefore finally extract the temperature of the scalar-tensor theory: looking at \eqref{consteq2}, we note that $h_{\mu\nu} \nabla^\nu T$ corresponds to the spatial projection of the temperature gradient. Then, since we are working in a comoving frame where the spatial temperature gradient vanishes, the temperature is related to the fluid acceleration, and \eqref{consteq2} is essentially a generalization of Fourier's law of thermal conduction, giving us
\begin{align}\label{tempdef}
K T= \frac{1}{\phi} \sqrt{-\nabla^\mu \phi \nabla_\mu \phi}.
\end{align}
The temperature of scalar-tensor gravity is therefore proportional to the  ``speed''  of the field, and vanishes when the scalar field stops evolving, which corresponds to GR. We can also see that the decomposition means that there is no viscous pressure in the case of $T^{(\phi)}_{\mu\nu}$, which means that $\zeta = 0$. Finally, the viscosity of the effective fluid is
\begin{align}
\eta = \frac{\sqrt{-\nabla^\mu \phi \nabla_\mu \phi}}{2\phi} .
\end{align}

The relaxation to GR (which corresponds to the equilibrium) can be seen by differentiating $KT$ with respect to the proper time along flow lines $\tau$. This yields the analogy of a \emph{thermal attractor} \cite{Faraoni:2025alq}:
\begin{align}
\frac{d (K T)}{d \tau} = 8 \pi (K T)^2 - \theta K T + \frac{\Box \phi}{\sqrt{-\nabla^\mu \phi \nabla_\mu \phi}}.
\end{align}
In particular, whether gravity is ``cooled'' or ``heated'' by this definition depends on the matter sector, which enters through the equations of motion of the field \eqref{eq: scalar field equation}. Therefore, the relaxation towards GR is recovered from a thermodynamics point of view \cite{Giardino:2022sdv}.

It is helpful to look at how the above formalism must be modified in the case of a minimally coupled scalar field~\cite{Faraoni:2022gry}. Indeed, using  the first law of thermodynamics $dU = - p \, dV  + T \, dS$, we can then consider a fixed particle number $N$ in a volume element and the respective number density $n$. Then, using the definitions $U = \rho N/n$, $V = N/n$, and $S = s N/n$, the first law becomes
\begin{align}
d \rho = \frac{\rho + p}{n} dn + T ds.
\end{align}
From this, we can therefore interpret the prefactor of $dn$ as a chemical potential, corresponding to the energy required to inject an additional particle: 
\begin{align}\label{chempotential_def}
\mu  =  \frac{\rho + p}{n}.
\end{align}
At the same time, the temperature cannot be specified without further constraining the system under consideration. Therefore, we specify the following minimal action:
\begin{align}\label{minimalact}
S = \frac{1}{2}\int d^4 x \, \sqrt{-g} \left[ R  + 2 {\cal L} (\phi, X) \right]+ S^{\rm (m)},
\end{align}
where $X \equiv  -\frac{1}{2}\nabla^\mu \phi \nabla_\mu \phi $.
In the Eckart frame (which coincides with the Landau frame since energy flow is given exactly by the particle flow), the energy density, pressure, and particle number density can be identified as~\cite{Piattella:2013wpa}
\begin{align}\label{thermdef1}
\rho  &= 2 X {\cal L}_{,X} - {\cal L}
\\
\label{thermdef2}
p &= {\cal L} 
\\
\label{thermdef3}
n &= \sqrt{2X} {\cal L}_{,X}.  
\end{align}
The chemical potential through \eqref{chempotential_def} is therefore 
\begin{align}\label{mudef}
\mu = \sqrt{2X}.
\end{align}
We note that the chemical potential depends solely on $X$: the only dependence on the field $\phi$ is implicit. 

We might expect that for a minimal action, the temperature can be defined in an analogous way, alongside the chemical potential $\mu$. However, attempting to do so leads us to a contradiction: minimal actions admit a perfect, non-dissipative fluid, and the existence of a spatial temperature gradient that corresponds to a dissipative heat current, which directly contradicts the perfect fluid description that features no such current. Definitions of temperature and chemical potential appear in \cite{Piattella:2013wpa}
but as noted in \cite{Faraoni:2022gry}, they are problematic since $T$ can be negative. The solution the authors propose is to switch between a thermal and dissipative picture depending on whether the coupling is nonminimal or minimal. 

Moving to a dissipative interpretation, we now seek a \emph{post hoc} relation analogous to~\eqref{consteq2} that the chemical potential $\mu$ satisfies. The relevant hydrodynamical law is Fick's diffusion law, whose relativistic form is \cite{Kremer:2013fxa}:
\begin{align}\label{fickslaw}
Q_\rho = - D (h_{\rho\sigma} \nabla^\sigma \mu + \mu {a}_\rho),
\end{align}
where $D$ corresponds to the diffusivity of the medium and  $Q_\mu$ denotes the diffusive flux density of particles, which follows from decomposing the particle number current as $N^\mu \equiv  n u^\mu + Q^\mu$. Indeed, we find that this law is satisfied in the case of the fluid admitted by a minimal action, as long as we take $Q_\mu = 0$.  This is exactly as we would expect for a non-dissipative fluid. Although it may seem paradoxical to trade the thermal picture for a dissipative picture even though we are in the minimal case which admits a non-dissipative fluid, Fick's law is not trivial because the spacetime dependence of $\mu$ is such that the two terms on the right-hand side of \eqref{fickslaw} vanish. In fact, the spatial derivative of $\mu$ yields 
\begin{align}
h_{\rho\sigma} \nabla^\sigma \mu = \frac{\nabla_\rho X}{\sqrt{2X}} + \frac{(\nabla_\sigma \phi) (\nabla^\sigma X) }{(2X)^{3/2}} \nabla_\rho \phi.
\end{align}
Then, using the expression for the acceleration \eqref{acceldef}, we find that $Q_\mu$ vanishes identically in the comoving frame \cite{Faraoni:2022gry}. This also means that the diffusivity $D$ is relegated to a dimensionful bookkeeping constant that can be absorbed into the definition of $\mu$.

We can finally differentiate the chemical potential with respect to the proper time $\tau$ which yields
\begin{align}\label{chem1}
\frac{d\mu}{d\tau} \equiv {u}^\rho \nabla_\rho \mu
\end{align}
and returns
\begin{align}\label{chem2}
\frac{d\mu}{d\tau}  =\frac{\nabla^\mu \phi \nabla_\mu X }{2 {X}}.
\end{align}
Through the expansion scalar $\theta = \nabla_\mu {u}^\mu$, we finally write: 
\begin{align}\label{chem3}
\frac{d\mu}{d\tau} = -\mu \theta + \Box \phi.
\end{align}
This equation controls whether GR is approached or not. The evolution of $\Box \phi$ depends on the matter as well, but generally, we can see that $\mu = 0 $ corresponds to diffusive equilibrium, since it also ensures $\Box\phi = 0$. Thus, instead of gravity being ``cooled'' to GR or ``heated'' away from it, it ``dilutes'' towards GR and ``condenses'' away from it: indeed, as expected, an infinitely diluted theory means that there is no flow of particles and we find ourselves in the~GR regime.

Therefore, the approach favored by many authors is that when dealing with nonminimally coupled theories, the thermodynamics of the theory is best captured by a picture with nonzero temperature and that relaxation towards GR involves $T\to 0$, whereas in minimally coupled theories, we must switch to a perfect fluid description, and GR appears not as a thermal equilibrium but rather as a diffusive equilibrium, where the chemical potential vanishes. However, this resolution does not contend with the issues of conformal dependence that appear in the definition of temperature as well as other thermodynamical quantities, which we will discuss in the next section.

\section{Invariant formulation of scalar-tensor theories}\label{sec:invariantscalartensor}

The diffusive equilibrium approach resolves the apparent contradictions of the thermal approach to minimal scalar-tensor theories. However, the thermal approach to nonminimal theories seems to be conformal frame-dependent, which poses a problem if we are to define parametrization-independent quantities. We recall that any general scalar-tensor theory coupled to matter, specified by four scalar ``model parameters'', can be written in its most general form as~\cite{Flanagan:2004bz,Jarv:2014hma},
\begin{align}\label{scalartensoraction}
S_{ST} = \frac{1}{2} \int d^4 x \,\sqrt{-g}  \big[ A(\phi) R - B(\phi) (\nabla\phi)^2  -2 V(\phi) \big] + S^{\rm (m) } [ e^{2\sigma(\phi)} g_{\mu\nu}, \chi]. 
\end{align}
Such an action captures the entire class of scalar-tensor theories coupled to matter: it, for instance, includes all Brans--Dicke theories. From now on, we will ignore the matter sector for simplicity, since the fluid description of the scalar-tensor sector and the effective stress-energy tensor does not explicitly depend on the matter coupling.

It is possible to move between representations (frames) of this theory via a frame transformation, which is simply a field-dependent rescaling of the metric followed by a field reparametrization:
\begin{align}
g_{\mu\nu} &\mapsto \hat g_{\mu\nu} = \Omega(\phi)^2 g_{\mu\nu},
\\
\phi &\mapsto \hat \phi = f(\phi).
\end{align}
The theory is then transformed to
\begin{align}\label{scalar_tensor_gen}
S_{ST} = \frac{1}{2} \int d^4 x \,\sqrt{-\hat g}  \big[ {\hat A} \hat R - \hat g^{\mu\nu} {\hat B} (\partial_\mu\hat\phi) (\partial_\nu \hat \phi)   -2 {\hat V} \big] 
\end{align}
where we have discarded a surface term and $\hat R$ is simply the scalar curvature defined in terms of $\hat g_{\mu\nu}$. The new model functions are given by 
\begin{align}
\hat A   &= \Omega^{-2} A,
\\
\label{Btransform}
\hat B  &= \Omega^{-2} \big[   f'^{-2} B   - 6 A  (\ln \Omega)'^2 + 6 A' (\ln \Omega)' \big],
\\
\hat V &= \Omega^{-4} V,
\end{align}
where we have suppressed arguments of $\phi$ and $\hat\phi$: it is understood that hatted functions take $\hat\phi$ as their argument through $\phi(\hat\phi)$ (which notably presupposes that the field redefinition is invertible). Primes indicate differentiation with respect to $\phi$. Expressing the model functions in terms of $\hat \phi = f^{-1}(\phi)$ further requires the use of the chain rule.

The transformed theory is known to be observationally indistinguishable from the original theory in the absence of a boundary. This means that, at the classical level at least, there is no preferred frame: any picture, regardless of whether matter is minimally coupled and the scalar field is nonminimally coupled, or whether matter is nonminimally coupled and the scalar field is minimally coupled, or even both are nonminimally coupled, must contain the same physics.  

There is a deep reason for the fact that changing the frame of the theory does not change the physics, related to the isomorphism between frame transformations and unit transformations \cite{Karamitsos:2025yyv}, along with the result that unit transformations do not change the physical content of a theory. From the Buckingham-$\pi$ theorem, we know it is possible to express any physically meaningful relation in terms of dimensionless quantities $\pi_i$  as
\beq
f(\pi_1, \ldots, \pi_n) = 0.
\eeq
By extension, given that the action is foundational in expressing the physical content of a theory, more so than the equations of motion (consider e.g. the path integral formulation and the various off-shell quantities which become important in QFT), we expect that we should be able to write it in a frame invariant manner. The process is analogous to nondimensionalization: instead of identifying independent dimensionless quantities $\pi_i$, we identify building blocks that are invariant under conformal transformations and transform as scalars under field reparametrizations. For a scalar-tensor theory given in \eqref{scalar_tensor_gen}, we may identify various invariants. However, much like a basis of a vector space, only two independent ones constructed solely from model functions emerge \cite{Jarv:2014hma,Jarv:2015kga,Jarv:2016sow}:
\begin{align}\label{invarPhi}
\Phi &\equiv \pm \int d \phi \, \sqrt{ \frac{B(\phi)}{A(\phi)} + \frac{3}{2} \left[\frac{A'(\phi)}{A(\phi)}\right]^2 },
\\
\label{invarU}
\mathfrak{U}(\Phi) &\equiv \frac{V(\phi(\Phi) )}{A(\phi(\Phi))^2},
\end{align}
where we note that writing $\phi(\Phi)$ requires us to invert $\Phi(\phi)$ (in certain situations, such as in the presence of poles, $\Phi(\phi)$ may not be invertible, which leads to a ``splitting'' of the original theory into multiple canonical theories, disjoint from one another \cite{Karamitsos:2019vor}).

In addition, we may define the invariant line element $d\mathfrak{s}$:
\begin{align}
d\mathfrak{s}^2 &\equiv A \,  g_{\mu\nu} \, dx^\mu dx^\nu 
\end{align}
as well as the invariant metric
\begin{align}\label{eq:invariant metric}
\mathfrak{g}_{\mu\nu} &\equiv A g_{\mu\nu}.
\end{align}
The invariant line element is arbitrary up to a multiplicative constant. However, unlike the metric in any particular frame, the invariant line element can be used to compare distances in any frame regardless of the spacetime-dependent system of units that is admitted by the choice of a non-constant nonminimal coupling function, which acts as a ruler. Indeed, tuning $A$ serves to change the measure of a dimensionful quantity (length): however, the corresponding nondimensionalized quantity (invariant length) does not change as we go from frame to frame.

A useful notion closely related to the frame invariant quantities is that of frame \emph{covariant} quantities~\cite{Burns:2016ric, Karamitsos:2017elm, Finn:2019aip, Finn:2020nvn}. While quantities such as the kinetic coupling $B(\phi)$ are neither invariant nor do they transform in a covariant manner, there are certain quantities that transform as densities under a frame transformation, which is to say, they pick up a conformal factor raised to some power. A quantity $Y$ is therefore said to be \emph{frame-covariant} if it transforms as 
\begin{align}
Y \mapsto \Omega^{-c_Y} Y,
\end{align}
with conformal weight $c_Y$ and we have suppressed any tensor indices of $Y$.  Such quantities are analogous to dimensionful quantities that pick up a scale factor under unit transformations, except here this scale factor is implicitly spacetime-dependent through its dependence on the scalar field. 

Frame-covariant quantities can of course be multiplied and divided to arrive at invariant quantities. However, we can readily see that the derivative of such a quantity is not frame-covariant: 
\beq
\partial_\mu Y \mapsto \Omega^{-c_Y} \partial_\mu Y +  Y \partial_\mu\Omega^{-c_Y},
\eeq
since it does not pick up an overall coefficient of a power of the conformal factor.
This is not a problem if we wish to work solely with invariant quantities, but having a frame-covariant derivative can be helpful in constructing invariant quantities by promoting partial derivatives to covariant derivatives (much like symmetries are gauged). This motivates the definition of a frame-covariant derivative~\cite{Postma:2014vaa,Burns:2016ric, Karamitsos:2017elm} that commutes with the conformal factor. Using the nonminimal coupling as analogous to a gauge (since it cannot be measured and it effectively amounts to a \emph{choice} that sets the system of units), we define the conformally covariant derivative as 
\begin{align}\label{conformalderiv}
D_{\mu} Y \equiv \partial_\mu Y  - \frac{c_Y}{2} \frac{\partial_\mu A }{A} Y.
\end{align}
Now the quantity $D_{\mu} Y$ transforms covariantly under a conformal transformation. We may also define a ``fully'' covariant derivative that is both conformally and gauge-covariant as
\begin{align}\label{fullcovder}
{\cal D}_{\mu} Y^\rho \equiv \nabla_\mu Y^\rho  - \frac{c_Y}{2} \frac{\partial_\mu A}{A} Y^\rho,
\end{align}
where the derivative now also respects the tensor structure of $Y^\rho$.

The definition of frame-covariant derivatives is useful because it gives us a recipe to construct invariant quantities even when derivatives are involved, and helps us keep track of conformal weights. Nonetheless, we note again this derivative is not strictly necessary: when working at the level of the action, we expect to always be able to write it in a manifestly invariant form, and taking derivatives of invariant quantities with respect to other invariant quantities will always lead to manifestly invariant equations of motion. Therefore, expressing everything in terms of the invariants~\eqref{invarPhi} and \eqref{invarU} suffices to have manifest frame invariance. Still, the conformal derivative, reminiscent as it is to a gauge-covariant derivative, is a straightforward way to promote quantities to their covariant counterparts. For instance, we can define a conformally covariant version of the Ricci scalar by promoting all partial derivatives to conformally-covariant derivatives \eqref{conformalderiv}
in the definition of the Christoffel symbols:
\begin{align}
\digamma^\rho_{\mu\nu} \equiv \frac{g^{\rho \sigma} }{2}(D_\mu g_{\sigma\nu} + D_\nu g_{\mu\sigma}  -D_\sigma g_{\mu\nu} ).
\end{align}
This results in the frame-covariant (and in particular, \emph{invariant}) Christoffel symbols, which can in turn be used to define the frame-covariant counterparts of the Riemann tensor, the Ricci tensor, and the Ricci scalar in the usual manner:
\begin{align}
  \mathfrak{R}^{\rho}{}_{\sigma\mu\nu} &=
   D_{\mu}\digamma^{\rho}{}_{\nu\sigma} -
  D_{\nu}\digamma^{\rho}{}_{\mu\sigma} +
  \digamma^{\rho}{}_{\mu\lambda}\digamma^{\lambda}{}_{\nu\sigma} -
  \digamma^{\rho}{}_{\nu\lambda}\digamma^{\lambda}{}_{\mu\sigma},
  \\
  \mathfrak{R}_{\mu\nu} &=  \mathfrak{R}^{\rho}{}_{\mu\rho\nu},
  \\
    \mathcal{R}  &= g^{\mu\nu}   \mathfrak{R}_{\mu\nu}.
\end{align}
Note that the Riemann and Ricci tensors defined this way are invariant, whereas the Ricci scalar is covariant: it transforms with weight $c_{\mathcal{R}} =2$. The invariant Ricci scalar, which is defined in terms of $\mathfrak{g}_{\mu\nu}$, is simply given as
\begin{align}
\mathfrak{R} = A(\phi)^{-2} \mathcal{R}.
\end{align}
In fact, by using the definition of the frame covariant derivative (with the appropriate weights), the invariant Ricci scalar can be written in terms of the standard Ricci tensor as:
\begin{align}
\mathfrak{R} = A^{-1}\big[R 
-  3A^{-1} g^{\mu\nu}  \nabla_\mu \nabla_\nu A  +  \frac{3}{2} A^{-2} g^{\mu\nu} (\partial_\mu A)(\partial_\nu A) \big],
\end{align}
which is formally identical to the Ricci scalar in the Einstein frame. Therefore, by putting all the invariant quantities together, we can now express the scalar-tensor action in a manifestly invariant form:
\begin{align}\label{scalar_tensor_invar} 
S = \frac{1}{2} \int d^4 x \,\sqrt{-\mathfrak{ g} }  \big[  \mathfrak{R} -  \mathfrak{g}^{\mu\nu } (\partial_\mu \Phi)  (\partial_\nu \Phi)   -2 \mathfrak{U} \big] 
\end{align}
where the invariant Ricci scalar $\mathfrak{R}$ is defined in terms of the invariant metric. We have left out the surface term but that too can be written solely in terms of invariant quantities \cite{Jarv:2014hma}.

The fact that we can write a scalar-tensor action in a frame invariant form is remarkable in its own right: it is an explicit demonstration that the frame in which a theory is expressed has no physical bearing at all on its physical content. This does not mean that non-invariant quantities are not helpful: they may be of use in calculations or in specifying a theory in a particular frame, which  is how most model building occurs in practice. However, we must stress a very important point: non-invariant quantities are not characteristic of a \emph{theory}. Rather, they are characteristic of a particular \emph{representation} of it, in this case, the particular frame in which it is expressed. This is a salient point because in order for thermodynamical quantities to truly be intrinsic to a theory, they must be invariant. However, as we shall see in the next section, some quantities as commonly defined in the literature are not invariant, which poses a challenge in our understanding of thermodynamics of the general class of scalar-tensor theories.
 
\section{Conformal dependence of thermodynamical quantities}
\label{sec:conformallydependentthermo}

As we discussed above, only frame invariant quantities can be said to be features of a theory rather than of its presentation. 
Indeed, a scalar-tensor theory can be viewed as an equivalence class of scalar-tensor actions, all of which share the same frame invariant quantities: the theory space is the action space quotiented out by frame transformations~\cite{Karamitsos:2025ugq}. In this sense, frame covariance can be viewed as analogous to general covariance. To understand this analogy, consider a point mass as it traverses spacetime. Its trajectory can be described without reference to a coordinate system and indeed exists independently of any coordinates within the manifold. Still, we can write down its equations of motion in some coordinate system, and we can transform to a different coordinate system while describing precisely the same trajectory. The same thing happens with a theory built out of gravitational and scalar degrees of freedom: the theory exists independent of our choice of parametrization (frame), and we can transform between different expressions of the theory (actions) while the physical information encoded by said expressions remains exactly the same. The rest frame is then analogous to the Einstein frame: in both cases, relative/frame-dependent quantities become equal to their absolute/invariant forms (e.g.\ transforming to the rest frame gives us the rest energy of a particle). Indeed, there is no experiment that can distinguish between the Jordan frame, i.e.\ measure the nonminimal coupling $A(\phi)$, much like there is no experiment that can measure absolute velocity. The analogy between frame covariance and general covariance is shown in Table~\ref{tab:covariance}.

\begin{table}
\centering
\begin{tabular}{l l}
\textbf{general covariance}  & \textbf{frame covariance} \\
\hline
trajectory   & theory    \\
observer/coordinate system    & frame \\
equations of motion    & action    \\
coordinate transformation    & frame transformation    \\
relative quantities    & frame-dependent quantities    \\
absolute quantities    & frame invariant quantities    \\
rest frame    & Einstein frame    \\
moving frame    & Jordan frame    \\
\end{tabular}
\caption{Correspondence between frame covariance and general covariance.}
\label{tab:covariance}
\end{table}

We can now examine whether temperature, as defined earlier, is frame invariant. First, we note that the definition in \eqref{tempdef} presupposes that we are working with a Brans--Dicke action. However, while every scalar-tensor theory can be cast in a Brans--Dicke form, the parametrization of $\phi$ is not unique. Performing a conformal transformation $g_{\mu\nu} \to \hat g_{\mu\nu}= \Omega(\phi)^2 g_{\mu\nu}$ on a Brans--Dicke theory returns:
\begin{align}
S_{\text{BD}} = \frac{1}{2}\int d^{4}x \sqrt{-  \hat g} 
\left[ 
(\Omega^{-2}\phi)   \hat R 
- \frac{\hat \omega(\phi) f'(\phi)^2}{\Omega^{-2}\phi}
{\hat g}^{\mu\nu} (\partial_\mu \phi)(\partial_\nu \phi) 
+ \ldots 
\right],
\end{align} 
where we omit the potential term (which does not enter explicitly into the temperature) and introduce an arbitrary function $f$ such that
\begin{align}
\frac{\hat \omega(\phi) f'(\phi)^2}{\Omega^{-2}\phi} =   \left(  \frac{\omega (\phi )}{\phi }
-\frac{6 }{\phi ^2 } \frac{\Omega '}{\Omega }
-6 \phi \frac{ \Omega'^2}{\Omega^2} \right).
\end{align}
We have the freedom to do this because a frame transformation is specified by two functions: the conformal factor and the field reparametrization, which we have not specified yet.
Then, if we identify $\hat \phi = f(\phi) \equiv \Omega(\phi)^{-2} \phi$, the action becomes
\begin{align}
S_{\text{BD}} = \frac{1}{2}\int d^{4}x \sqrt{-  \hat g} 
\left[ 
\hat \phi   \hat R 
- \frac{\hat \omega(\hat \phi)}{\hat \phi}
{\hat g}^{\mu\nu} (\partial_\mu \phi)(\partial_\nu \phi) 
+ \ldots
\right].
\end{align}
Analogously, the temperature \eqref{tempdef} of the theory transforms into
\begin{align} \label{temptrans}
K\hat T &= \frac{1}{\hat \phi} \sqrt{- (\hat \nabla^\mu \hat \phi)( \hat \nabla_\mu \hat \phi)}
\\
&= \frac{ \Omega (\phi) f'(\phi)}{\phi}  \sqrt{-  (\nabla^\mu   \phi)(  \nabla_\mu \phi)}
\\
&= \frac{\Omega (\phi )-2 \phi  \Omega '(\phi )}{ \Omega (\phi )^2} KT.
\end{align}
We can therefore immediately see that there is an explicit issue: the temperature can be arbitrarily tuned with a frame transformation. This is a fatal error: if temperature is to be a quantity intrinsic to a theory rather than its representation, then \eqref{tempdef} is not an appropriate definition since it depends on our choice of frame. In fact, ``making'' the temperature zero is equivalent to transforming to the Einstein frame, which we can see by solving for $\Omega(\phi)$ when setting $\hat T= 0$ in~\eqref{temptrans}.

The issue persists even if we move beyond Brans--Dicke theories. The definition of temperature for a more general scalar-tensor action suffers from the same frame dependence problem. The thermodynamics of modified gravity theories have been studied for the generalized Horndeski-type theories \cite{Giusti:2021sku, Gallerani:2024gdy}:
\begin{align}
S = \frac{1}{2} \int d^4 x \, \sqrt{-g} \, 
\Big[ 
G_2(\phi,X) 
-G_3 (\phi,X)  \Box\phi 
+G_4(\phi,X) R
\Big].
\end{align}
The effective four-velocity (the definition of which is unchanged from \eqref{fourvelocitydef}) leads to the following expression for the four-acceleration:
\begin{align}
a^\mu = -\frac{1}{2X} \left( \nabla^\mu X + \frac{\nabla^\rho X \nabla_\rho \phi}{2X} \nabla^\mu \phi\right)
\end{align} 
and the effective heat flux is found to be
\begin{align}
q_\mu = \frac{G_{4,X } - X G_{3,X}}{G_{4} \sqrt{2X}} \left( \nabla_\mu X + \frac{\nabla_\rho X \nabla^\rho \phi}{2X} \nabla_\mu \phi\right) .
\end{align}
Crucially, the generalized Fourier law \eqref{consteq2} is also unchanged, and thus the temperature can be extracted by once again identifying $K h_{\mu\nu} \nabla^\nu T = 0$ (since in the comoving frame the temperature gradient vanishes), returning
\begin{align}
K { T} = \frac{\sqrt{2X} (G_{4,\phi}- X G_{3,X})}{G_4}.
\end{align}
We will now restrict our attention to the scalar-tensor subclass of Horndeski theories (the full Horndeski theory is form invariant under disformal transformations, but this is beyond the scope of this paper).  Using the form of the action prescribed in \eqref{scalartensoraction},
we find the following expression for the ``temperature of gravity'':
\begin{align}\label{Tgeneraldef}
K T = \frac{A'(\phi) }{A(\phi)}\sqrt{-\nabla^\mu \phi \nabla_\mu \phi},
\end{align}
which clearly is a frame-dependent statement.\footnote{An alternative approach to temperature and conductivity, in which $KT$ is split such that $K = 1/G_4$ is not constant \cite{Giusti:2021sku} still has the same issue, since $T=1/A(\phi)$, which is still not a frame invariant expression.} Once again, this frame-dependence is not trivial, e.g.\ up to a constant rescaling. This means that it is again possible to tune the temperature by appropriately selecting a frame. 

A frame-dependent temperature is not necessarily a problem: after all, in relativistic thermodynamics, the transformation formulae for standard thermodynamical temperature are not widely agreed upon \cite{Farias:2017} and may depend on assumptions regarding internal heat flow \cite{Biro:2009ki}, but it is possible to define \emph{proper temperature} by looking at the rest frame of a moving body. This means that we could single out a particular frame in which we define temperature. The Einstein frame, uniquely defined as it is (up to an overall constant), is an ideal candidate: this also satisfies the rest frame--Einstein frame correspondence in Table~\ref{tab:covariance}. However, we quickly discover that in this case, the temperature is identically zero: the Einstein frame is minimally coupled, and as a result, admits a perfect fluid. 

Taking the frame invariant action \eqref{scalar_tensor_invar} as our starting point and working solely with frame invariant quantities, we arrive at the same result: as the invariant action is minimally coupled, the resulting temperature also vanishes. The same result can still be obtained if we take the definition of temperature in  \eqref{Tgeneraldef} and promote the field derivative to a covariant derivative: since $D_\mu A = 0$,\footnote{This is the mathematical expression of the fact that we cannot measure changes in $A(\phi)$. This is because $A(\phi)$ is not a feature of the theory: it is merely an artifact of the frame we choose to express it in. Indeed, for a scalar-tensor theory, there is no experiment that can reconstruct the nonminimal coupling function, much like there is no experiment that measures gauge fields (only gauge-invariant quantities e.g.\ the phase difference in the Aharonov--Bohm effect).} the temperature once again vanishes. 

It then appears that the notion of proper temperature for scalar-tensor theories is trivial, much like the notion of ``proper velocity'': both are by definition zero, and both can be arbitrarily tuned by selecting an appropriate reference frame. A vanishing temperature for scalar-tensor gravity is consistent with the Einstein frame treatment that trades the temperature for a chemical potential. This indicates that the frame invariant effective fluid description of scalar-tensor gravity is not thermal, but rather diffusive: the imperfect fluid description with a non-zero temperature is merely an artifact of working in a nonminimal frame. We develop the frame invariant diffusive analogue of scalar-tensor theories in the next section.
 
\section{Frame invariant diffusive equilibrium}\label{sec:frameinvariantequilibrium}

In order to build a frame invariant notion of diffusive equilibrium, we begin from the fundamental kinematic quantities using the frame-covariant formalism we have motivated. We first note that the fluid velocity transforms as follows under a frame transformation:
\begin{align}
u_\mu &\mapsto  \Omega  u_\mu,
\\
u^\mu  &\mapsto \Omega^{-1} u^\mu.
\end{align}
This immediately tells us that the four-acceleration is \emph{not}  covariant, as it picks up additional terms:
\begin{align}
a^\mu \mapsto \ \Omega^{-2} u^\nu ( \nabla_\nu  u^\mu  -  u^\mu\nabla_\nu \ln \Omega ),
\end{align}
and as a result, the imperfect fluid decomposition is not invariant. However, if we start from the invariant metric \eqref{eq:invariant metric}, we can decompose it as
\begin{align}
\mathfrak{ h}_{\mu\nu}  \equiv \mathfrak{g}_{\mu\nu}+ \mathfrak{u}_\mu \mathfrak{u}_\nu,
\end{align}
where we now have the invariant induced metric. We can use this to define the invariant four-acceleration:
\begin{align}
\mathfrak {a}^\mu \equiv  \mathfrak{ u}^\nu  \nabla^{(\mathfrak{g})}_\nu \mathfrak{ u}^\mu,
\end{align}
where the derivative $\nabla^{(\mathfrak{g})}_\nu$ is compatible with $\mathfrak{g}_{\mu\nu}$. 
This definition can also be motivated by promoting  $\nabla_\mu$ to its fully covariant counterpart $\mathcal{D}_\mu$ given in \eqref{fullcovder}:
\begin{align}
\mathfrak{a}^\mu \equiv u^\nu {\cal D}_\nu u^\mu.
\end{align}
Thanks to the transformation properties of $u_\mu$ and $u^\mu$, the two definitions are equivalent. 

We can explicitly write the invariant four-velocity as
\begin{align}
\mathfrak{ u}^\mu \equiv \frac{{\partial}^\mu \Phi }{\sqrt{- \mathfrak{g}^{\rho\sigma}\partial_\rho \Phi \partial_\sigma \Phi}},
\end{align}
Looking back at the minimal action \eqref{minimalact}, of which the invariant action \eqref{scalar_tensor_invar} is a special case, we can straightforwardly write down the frame invariant chemical potential through \eqref{mudef}:
\begin{align}\label{eq:MfromX}
\mathfrak{M} \equiv \sqrt{2 \mathfrak{X} },
\end{align}
where $\mathfrak{X}$ is simply the frame invariant kinetic term:
\begin{align}
\mathfrak{X} \equiv -\frac{1}{2} \partial_\mu \Phi \partial^\mu \Phi.
\end{align}
This means that 
Fick's diffusion law \eqref{fickslaw} written in a frame invariant form is satisfied:
\begin{align}\label{eq: Fick invariant}
\mathfrak{Q}^\mu = - D (\mathfrak{h}^{\mu\nu} \nabla^{(\mathfrak{g})}_\nu \mathfrak{M} + \mathfrak{M} \mathfrak{a}_\mu),
\end{align}
where $\mathfrak{Q}^\mu$ is the frame invariant counterpart of the diffusive flux density of particles, which follows from the decomposition of the (invariant) particle number current as $\mathfrak{N}^\mu = \mathfrak{n} \mathfrak{u}^\mu + \mathfrak{Q}^\mu$. Once again, similar to the minimal picture, the current is purely convective: $\mathfrak{Q}^\mu$ vanishes identically, and the frame invariant version of Fick's law is satisfied, precisely because the invariant chemical potential $\mathfrak{M}$ is not constant. The result is that in the comoving frame, we still observe no dissipative particle flow.

We can thus finally write the frame invariant chemical potential explicitly for \emph{any} scalar-tensor theory given in \eqref{scalartensoraction}
\begin{align}
\mathfrak{M}  \equiv  \sqrt{-  \left[B(\phi)  + \frac{3}{2}  \frac{A'(\phi)^2}{A(\phi)}\right]  \, g^{\mu\nu}  \, (\partial_\mu \phi) \, (\partial_\nu \phi)} \, .
\end{align}
This frame invariant chemical potential then becomes the truly relevant quantity for a nonminimal theory that can be cast in an invariant manner: the frame invariant temperature vanishes, and does not enter the thermodynamical picture.

We note that as a system with zero temperature, its thermodynamics (as opposed to the hydrodynamics) are strictly speaking trivial. The invariant density $\varrho$, invariant pressure $\mathfrak{p}$, and particle number density $\mathfrak{n}$ are defined simply by applying \eqref{thermdef1}, \eqref{thermdef2}, and \eqref{thermdef3} to the invariant action \eqref{scalar_tensor_invar}, which returns
\begin{align}
\varrho  &= \mathfrak{X} + \mathfrak{U}
\\
\mathfrak{p} &= \mathfrak{X} - \mathfrak{U}
\\
\mathfrak{n} &=   \sqrt{2  \mathfrak{X}}.  
\end{align}
We note that the invariant particle number density in this case coincides with the invariant chemical potential, $\mathfrak{M} = \mathfrak{n}$.  

It is important here to make a comment regarding entropy. Traditionally, entropy for a minimally coupled theory is identified as $\phi/n$, but we can see that this definition is problematic as it transforms in a highly nonlinear fashion under a field reparametrization. Furthermore, the fact that temperature vanishes means that the definition of entropy itself is problematic. The invariant first law of thermodynamics becomes
\begin{align}
d \varrho = \frac{\varrho + \mathfrak{p}}{\mathfrak{n}} d\mathfrak{n} + \mathfrak{T} d\mathfrak{s},
\end{align}
where we have formally introduced the ``invariant temperature'' $\mathfrak{T}$ and ``invariant entropy density''  $\mathfrak{s}$. However, since we know that $\mathfrak{T}=0$, entropy becomes indeterminate: any variation will satisfy the first law of thermodynamics.

Finally, we can see that the scalar field equations of motion take on the generalized Klein--Gordon form
\begin{align}
 \Box^{(\mathfrak{g})} \Phi + \mathfrak{U}_{, \Phi} = 0.
\end{align}
where we have defined $\Box^{(\mathfrak{g})} \equiv\mathfrak{g}^{\mu\nu} \nabla^{(\mathfrak{g})}_\mu \nabla^{(\mathfrak{g})}_\nu$. 

This equation can be written in the form of the invariant continuity equation. First we write down the frame invariant energy-momentum tensor:
\begin{align}
\mathfrak{T}_{\mu\nu} &= (\partial_\mu \Phi)( \partial_\nu \Phi) - \frac{\mathfrak{g}_{\mu\nu} \mathfrak{g}^{\rho\sigma}}{2} (\partial_\rho \Phi)( \partial_\sigma \Phi) - \mathfrak{U} \mathfrak{g}_{\mu\nu} \, ,
\\
&= (\varrho + \mathfrak{p}) \mathfrak{u}_\mu \mathfrak{u}_\nu +  \mathfrak{p}\mathfrak{g}_{\mu\nu},
\\
&= \varrho \mathfrak{u}_\mu \mathfrak{u}_\nu + \mathfrak{p} \mathfrak{h}_{\mu\nu},
\end{align}
We can take the covariant derivative: 
\begin{align}
\nabla^{(\mathfrak{g})}_\mu \mathfrak{T}^{\mu\nu} = \mathfrak{u}^\nu \left[\frac{d\varrho}{d\mathfrak{t}} + \frac{d\mathfrak{p}}{d\mathfrak{t}}  +  (\varrho +\mathfrak{p})\Theta\right] + \nabla^{(\mathfrak{g}) \nu} \,\mathfrak{p} + (\varrho + \mathfrak{p}) \mathfrak{a}^\nu,
\end{align}
where $\mathfrak{t}$ is the invariant proper time, and is defined analogously to $\tau$ as in \eqref{fouraccelerationdef}:
\begin{align}
\mathfrak{a}^\mu \equiv \frac{d \mathfrak{u}^\mu}{d\mathfrak{t}} = \mathfrak{u}^\nu \nabla_\nu \mathfrak{u}^\mu.
\end{align}
Since $\mathfrak{u}^\nu \mathfrak{u}_\nu =-1$ and $\mathfrak{u}^\mu \mathfrak{a}_\mu =0$, contracting with $\mathfrak{u}_\nu$ returns
\begin{align}
\frac{d \varrho}{d\mathfrak{t}} + (\varrho + \mathfrak{p})\Theta = 0,
\end{align}
which is indeed the continuity equation: it takes the standard form in the Einstein frame and in special cases for spacetime such as FLRW. We can also write it in terms of the invariant field:
\begin{align}\label{fieldeq}
\frac{ d^2 \Phi}{d\mathfrak{t}^2} + \Theta \frac{ d  \Phi}{d\mathfrak{t} }  + \mathfrak{U}_{, \Phi} = 0,
\end{align}
which again mimics the usual field equation in expanding spacetime.

Finally, we can examine the equation that governs the evolution of the chemical potential along the lines of  \eqref{chem1}-\eqref{chem3}. We differentiate it with respect to the frame invariant proper time~$\mathfrak{t}$ along the flow lines to obtain:
\begin{align}
\frac{d\mathfrak{M}}{d\mathfrak{t}} &\equiv \mathfrak{u}^\mu \partial_\mu \mathfrak{M} 
= \frac{\mathfrak{g}^{\mu\nu}(\partial_\mu \Phi)( \partial_\nu \mathfrak{X})}{2 \mathfrak{X}}.
\end{align}
This gives
\begin{align}
\frac{d\mathfrak{M}}{d\mathfrak{t}} = -\mathfrak{M} \Theta + \Box^{(\mathfrak{g})} \Phi,
\end{align}
where we have straightforwardly extended the definition of the invariant expansion scalar $\Theta \equiv \nabla^{(\mathfrak{g})}_\mu \mathfrak{u}^\mu$ from the usual definition. 
This equation, much like \eqref{chem3}, is an identity which on its own does not tell us anything about how the chemical potential evolves. However, if endowed with additional information, in this case the d'Alembertian of the field $ \Box^{(\mathfrak{g})} \Phi$, it will tell us whether the chemical potential (and also the particle density, since $\mathfrak{M} = \mathfrak{n}$ identically) dilutes or condenses. In the absence of matter, we can straightforwardly use \eqref{fieldeq} to write
\begin{align}
\frac{d\mathfrak{M}}{d\mathfrak{t}} = -\mathfrak{M} \Theta - \mathfrak{U}_{,\Phi}.
\end{align}
For a vanishing potential, we can straightforwardly solve for $\mathfrak{M}$:
\begin{align}
\mathfrak{M}(\mathfrak{t}) = \mathfrak{M}_0 \exp \left[-\int_{\mathfrak{t}_0}^{\mathfrak{t}}  d\mathfrak{t}' \Theta\right],
\end{align}
which, for a monotonically increasing or decreasing $\Theta(\mathfrak{t})$, implies that GR is either asymptotically recovered or that the theory diverges from GR, respectively.

In the end, the invariant chemical potential controls dilution towards GR or condensation away from it, and GR is unambiguously and frame-agnostically specified by $\mathfrak{M}= 0$. Therefore, the frame invariant thermodynamical picture of \emph{all} scalar-tensor theories, regardless of minimal or nonminimal coupling, is very simple: temperature is always zero, and a non-uniform chemical potential controls the departure from GR. At the same time, this chemical potential is such that there is no diffusive current as expected of a perfect fluid for $\Phi$ even if $\mathfrak{M}$ is not constant. 

We conclude by presenting a summary of the link between the invariant field quantities to the hydrodynamical and thermodynamical counterparts in Table~\ref{tab:fluidcorrespondence}. These quantities are formally similar to the Einstein frame quantities, but they are completely independent of our choice of frame, as each term is individually conformally- and field reparametrization-invariant. Temperature is special because it is invariant in a trivial sense: it vanishes entirely regardless of the underlying theory which renders entropy indeterminate.
\begin{table}
\centering
\begin{tabular}{l l}
\textbf{fluid quantity}  & \textbf{frame invariant quantity} \\
\hline  \\[1pt]
energy density $\rho$                 & $\mathfrak{X} + \mathfrak{U}$      \\[6pt]  
pressure    $p$             & $\mathfrak{X} - \mathfrak{U}$ \\[6pt]
number density  $n$                & $\sqrt{2\mathfrak{X}}$    \\[6pt]
chemical potential $\mu$              &  $\sqrt{2\mathfrak{X}}$    \\[6pt]
temperature   $T$                & $0$ \\[6pt]
entropy density $s$                    & indeterminate    \\[6pt]
\end{tabular}
\caption{Correspondence between fluid quantities and their frame invariant counterparts.}
\label{tab:fluidcorrespondence}
\end{table}

\section{Single-component fluid from multiple fields}\label{sec:multifield} 

So far, we have demonstrated that a single-field action can be described invariantly in terms of hydrodynamical quantities. As a result, we may expect that for multiple fields, the action admits a multi-component fluid. However, as we will see in this section, this interpretation involves a particular subtlety regarding the parametrization of the field in the Einstein frame.

We consider the general multifield action in the Einstein frame without loss of generality \cite{Kuusk:2015dda,Karamitsos:2017elm}:
\begin{align}\label{scalar_tensor_invar_multifield} 
S = \frac{1}{2} \int d^4 x \,\sqrt{-\mathfrak{ g} }  \big[  \mathfrak{R} -  \mathfrak{g}^{\mu\nu } \mathfrak{G}_{IJ}(\partial_\mu \phi^I)  (\partial_\nu \phi^J)   -2 \mathfrak{U} \big] ,
\end{align}
where the field-space metric $\mathfrak{G}_{IJ}$ is defined as
\begin{align}
\mathfrak{G}_{IJ} = \frac{B_{IJ}}{A} + \frac{3}{2} \frac{A_{,I} A_{,J}}{A^2},
\end{align}
where $B_{IJ}$ is the kinetic prefactor in the original (non-explicitly invariant) form of the action. As expected, this quantity transforms as a tensor under field reparametrizations, picking up Jacobian factors, and is invariant under conformal transformations.

Using the field space metric can help us define the single-field equivalent as 
\begin{align}
d\Phi^2 = \mathfrak{G}_{IJ} d\phi^I d\phi^J.
\end{align}
With this definition, the invariant multifield action is formally identical to the invariant single-field action \eqref{scalar_tensor_invar}, and the chemical potential \eqref{eq:MfromX} remains unchanged:
\begin{align}
\mathfrak{M} &\equiv \sqrt{2\mathfrak{X}}
\\
&=\sqrt{-\mathfrak{g}^{\mu\nu} {\mathfrak G}_{IJ} \, \partial_\mu \phi^I \, \partial_\nu \phi^J},
\\
&= \sqrt{-\mathfrak{g}^{\mu\nu}   \partial_\mu \Phi \, \partial_\nu \Phi}.
\end{align}
Fick's law \eqref{eq: Fick invariant} then remains unchanged, and as usual, $\mathfrak{M} = 0$ means that the theory has relaxed to GR.

We can still define multiple four-velocities and multiple flux densities, all indexed by the species index, thus behaving as field-space vectors. We can begin by trying the following expression for the individual field velocities:
\begin{align}
\mathfrak{u}^I_\mu \equiv \frac{\nabla_\mu \phi^I}{\sqrt{-\mathfrak{g}^{\rho \sigma} \mathfrak{G}_{AB} (\partial_\rho \phi^A )(\partial_\sigma \phi^B)}},
\end{align}
which would ensure that $\mathfrak{u}^{\mu I} \mathfrak{u}_{\mu I} = -1$, since index gymnastics for the species indices are performed with the field-space metric $\mathfrak{G}_{IJ}$. This would allow a decomposition like 
\begin{align}
\mathfrak{g}_{\mu\nu} = \mathfrak{h}_{\mu\nu} + \mathfrak{G}_{IJ} \mathfrak{u}^I_\mu \mathfrak{u}^J_\nu,
\end{align}
which is still reparametrization invariant (in addition to being conformally invariant).

It is difficult, however, to see how we could define separate chemical potentials. Indeed, at first glance, $N$ fields seem like they would imply  that there should be $N$ chemical potentials, one for each ``species''. However, in this context, the notion of species is highly dependent on the parametrization of the action: the interaction between species depends on the coupling between the cross-terms, which become different under a reparametrization of the fields. Mathematically, it is not clear how we would be able to unambiguously get $N$ distinct expressions from the general kinetic term
\begin{align}
\mathfrak{X}^{IJ} = -\frac{\mathfrak{g}^{\mu\nu}}{2} \partial_\mu \phi^I \partial_\nu \phi^J,
\end{align}
which has more than $N$ components. 

We can examine the special case in which $\mathfrak{G}_{AB}$ happens to be Euclidean, which can also be achieved in the field-space analogue of the local inertial frame. The action then becomes
\begin{align}\label{scalar_tensor_invar_diag} 
S = \frac{1}{2} \int d^4 x \,\sqrt{-\mathfrak{ g} }  \Big[  \mathfrak{R} -   \mathfrak{g}^{\mu\nu } \sum_{I} \,  (\partial_\mu \phi^I)  (\partial_\nu \phi^I)  -2 \mathfrak{U} \Big].
\end{align}
Therefore, it appears that we may define the $N$ distinct chemical potentials as
\begin{align}
\mathfrak{M}_I \equiv \sqrt{ -  \mathfrak{g}^{\mu\nu } \partial_\mu \phi^I \partial_\nu\phi^I},
\end{align}
where the field space indices are not summed over. As a result, the single-field chemical potential appears simply as the Euclidean norm of the other potentials:
\begin{align}
\mathfrak{M} = \sqrt{\sum_I \mathfrak{M}_I^2}.
\end{align}
However, because of this identification, Fick's law \eqref{eq: Fick invariant} does not decompose linearly. 
Therefore, the multi-fluid analogy stops short of a system of linear equations describing $N$ particle species evolving according to a coupled set of diffusion equations, even if multiple conserved charges exist due to shift symmetry. As such, in the presence of multiple scalar fields, the natural way to proceed is to treat the fields as a single fluid condensing away from or diffusing to GR.

\section{Conclusions}\label{sec:conclusions}

In this work, we examined the thermodynamical interpretation of scalar-tensor theories of gravity. After reviewing the well-established imperfect fluid description of nonminimally coupled scalar fields in the Brans--Dicke picture, we demonstrated explicitly that the usual definition of the temperature of the fluid admitted by scalar-tensor theories is not a property of the underlying theory, but rather of its representation: it transforms nontrivially under conformal transformations and can be arbitrarily tuned by an appropriate choice of frame. This indicates that the thermodynamical picture commonly employed in interpreting nonminimal scalar-tensor theories does not capture invariant physical information. 

Arguing that physically meaningful quantities must be expressible in a frame invariant manner, we employed the frame invariant and frame-covariant formalism of scalar-tensor theories, demonstrating that the invariant action admits a perfect fluid. As a result, the invariant temperature vanishes identically. Instead, much like in minimal theories, the relevant invariant thermodynamical quantity is the chemical potential. However, this is a general feature of scalar-tensor theories. In this invariant picture, GR appears as the state of diffusive equilibrium rather than the state of thermal equilibrium: indeed, GR corresponds to a vanishing chemical potential. We further extended this invariant construction to multifield scalar-tensor theories, finding that despite the presence of multiple scalar degrees of freedom, the invariant formulation naturally admits a single effective fluid description. 

In the future, it will be interesting to consider   
how the invariant description can be generalized to more extensive scalar-tensor theories, such as the Galileon, Horndeski, and beyond (see \cite{Giusti:2021sku,Miranda:2022wkz,Faraoni:2023hwu,Miranda:2024dhw,Gallerani:2024gdy,Pereira:2025dmk}). Indeed, in such theories, a conformal transformation may eliminate the nonminimal coupling, but the noncanonical form of the kinetic term suggests that the diffusive picture of the relaxation to GR may be more complicated. Furthermore, we have not considered perturbations: it has been shown that the effective fluid decomposition of perturbations of scalar-tensor gravity in the Jordan frame also satisfy Eckart's first-order thermodynamics \cite{Pereira:2026xog}. It would be interesting to study whether frame invariant perturbations preserve the vanishing temperature, or whether a finite temperature can be meaningfully assigned to them. Finally, one may also try to realize the hydrodynamical and thermodynamical picture for scalar fields nonminimally coupled to non-Riemannian geometric models of gravity. While the invariant formulation of Palatini scalar-tensor gravity is quite straightforward \cite{Kozak:2018vlp,Jarv:2020qqm,Jarv:2024krk}, and in the metric and symmetric teleparallel versions of scalar-tensor gravity \cite{Hohmann:2018rwf,Jarv:2018bgs} the admission of conformal invariants introduces extra terms into the action \cite{Hohmann:2018ijr,Hohmann:2023olz}, the investigation of scalar fields nonminimally coupled to Einstein-Cartan, Weyl, or more general metric-affine gravity has barely begun (see e.g.\ \cite{Kiefer:2017nmo,Cid:2017wtf,Karananas:2021gco,Rigouzzo:2022yan}).

\section*{Acknowledgments}

This work was supported by the Estonian Research Council via the Center of Excellence ``Foundations of the Universe'' TK202U4. LJ was also supported by the  Estonian Research Council team grant PRG2608 and SK was supported by the Estonian Research Council Mobilitas 3.0 incoming postdoctoral grant MOB3JD1233 ``Inflationary Nonminimal Models: An Investigative Exploration''.

\bibliographystyle{utphys}
\bibliography{refs}

\end{document}